\documentclass[prb,aps]{revtex4}
\usepackage{epsfig}
\usepackage{graphicx}

\newcommand{\AmS}{{\protect\the\textfont2
  A\kern-.1667em\lower.5ex\hbox{M}\kern-.125emS}}
\newcommand{\vnabla}{{\mbox{\boldmath$\nabla$}}}
\newcommand{\veta}{{\mbox{\boldmath$\eta$}}}

\newcommand{\vR}{{\mbox{\boldmath$R$}}}

\newcommand{\vk}{{\mbox{\boldmath$k$}}}
\newcommand{\vv}{{\mbox{\boldmath$v$}}}

\newcommand{\vsk}{{\small \mbox{\boldmath$k$}}}

\newcommand{\hvk}{\hat{\mbox{\boldmath$k$}}}

\newcommand{\vd}{\mbox{\boldmath$d$}}

\newcommand{\mhz}{\hat{\mbox{\boldmath$z$}}}
\newcommand{\vh}{\mbox{\boldmath$h$}}
\newcommand{\vH}{\mbox{\boldmath$H$}}
\newcommand{\vS}{\mbox{\boldmath$S$}}
\begin{document}

\title{Quasi-classical determination of the in-plane magnetic field phase diagram of superconducting Sr$_2$RuO$_4$}
\author{R.P. Kaur $^1$, D.F. Agterberg $^1$, and  H. Kusunose $^2$}
\address{$^1$ Department of Physics, University of Wisconsin-Milwaukee, Milwaukee, WI 53211}
\address{$^2$ Department of Physics, Tohoku University, Sendai 980-8578, Japan}


\begin{abstract}
We have carried out a determination of the
magnetic-field-temperature (H-T) phase diagram for realistic
models of the high field superconducting state of tetragonal
Sr$_2$RuO$_4$ with fields oriented in the basal plane. This is
done by a variational solution of the Eilenberger equations.
This has been carried for spin-triplet gap functions with a
$\vd$-vector along the $c$-axis (the chiral $p$-wave state) and
with a $\vd$-vector that can rotate easily in the basal plane. We
find that, using gap functions that arise from a combination of
nearest and next nearest neighbor interactions, the upper critical
field can be approximately isotropic as the field is rotated in
the basal plane. For the chiral $\vd$-vector, we find that this
theory generically predicts an additional phase transition in the
vortex state. For a narrow range of parameters, the chiral
$\vd$-vector gives rise to a tetracritical point in the H-T phase
diagram. When this tetracritical point exists, the resulting phase
diagram closely resembles the experimentally measured phase
diagram for which two transitions are only observed in the high
field regime. For the freely rotating in-plane $\vd$-vector, we
also find that additional phase transition exists in the vortex
phase. However, this phase transition disappears as the in-plane
$\vd$-vector becomes weakly pinned along certain directions in the
basal plane.
\end{abstract}

\vspace{7 cm}

\maketitle
\section{Introduction}
It is widely believed that tetragonal Sr$_2$RuO$_4$
\cite{mae94,mae01,mac03} is a spin-triplet chiral $p$-wave
superconductor. In particular, a pairing state characterized by a
gap function $\vd=\hat{z}(k_x\pm i k_y )$ best explains existing
experimental results.  The observation of the appearance of local
magnetic moments below the superconducting transition temperature
by muon spin-resonance measurements of Luke {\it et al.}
($\mu$SR)\cite{luk98} can qualitatively be accounted for by the
two-fold degeneracy of the order parameter (a non-degenerate order
parameter cannot give rise to local magnetic moments). The nuclear
magnetic resonance\cite{ish98,mur04} and spin polarized neutron
scattering measurements\cite{duf00} carried out for fields applied
perpendicular to the four-fold symmetric $c$-axis show that the
spin susceptibility is unchanged by the normal to superconductor
transition. This is naturally explained by $\vd=\hat{z}(k_x\pm i
k_y )$ since $\vd$-vector is perpendicular to the magnetic field
for which no change of spin susceptibility is expected. Also, the
Josephson experiments of Liu {\it et al.} arguably place the
strongest constraint on the orientation of ${\vd}$-vector
\cite{nel04,jin00} and also implies a chiral $p$-wave
superconducting state with $\vd\parallel \hat{z}$. Finally, the
observed field distribution of the vortex lattice for the field
along the $c$-axis is not consistent with a non-degenerate (single
component) order parameter but can be accounted for by the chiral
$p$-wave state\cite{kea00,ris98}.

These experiments provide a convincing picture in favor of a
chiral $p$-wave superconductor. However, there are some
experiments that do not directly support this state. In
particular, the more recent Knight shift measurements of Murakawa
{\it et al.} have been carried out for the field along the
$c$-axis \cite{mur04}. These measurements reveal no change in the
spin susceptibility. For this field orientation, this would lead
to the conclusion that $\vd$-vector is in the basal plane, not
along the $c$-axis as would be the case for the chiral $p$-wave
state. The simplest interpretation of this experiment is that the
magnetic field is sufficiently strong as to rotate $\vd$ from
$\hat{z}$ to the basal plane. This would imply that the transition
temperatures for $\vd$ in the plane are close but slightly less
than that for $\vd$ along $\hat{z}$. This is possible if
spin-orbit coupling is weak. Another explanation for the Knight
shift data is that the $\vd$-vector is in the basal plane, but
free to rotate in the plane. This would require all four possible
in-plane degrees of freedom to be degenerate (or at least nearly
degenerate). If this is the case then for any in-plane field
orientation, $\vd$-vector will have an in-plane component
perpendicular to the field. Consequently, the spin susceptibility
will remain unchanged for
fields applied in the basal plane as well.


Another difficulty with the chiral $p$-wave state is that, for
magnetic fields applied in the basal plane, there are two
qualitative predictions for which there is little experimental
evidence. These are:  \\
(i) the existence of an anisotropy in the upper critical field as
the field is rotated perpendicular to the four-fold symmetry axis
that {\it does not vanish as} $T\rightarrow T_c$ \cite{gor87};\\
(ii) the existence of a phase transition in the vortex state in
addition to the usual transitions at $H_{c2}$ and $H_{c1}$. This
additional transition is due to a change in the
structure of the order parameter \cite{agt98}.

The primary goal of this work is to understand if there are
microscopic theories of the chiral $p$-wave state that  can lead
to situations where the above (i), (ii) predictions do not hold.
We find that it is plausible that one of the two predictions does
not hold, but less likely that both do not hold. Intriguingly,
this analysis also points to the possibility of a tetra-critical
point in the H-T phase diagram. This tetracritical point has
features that agree with recent experimental measurements in high
magnetic fields \cite{mao00}. Given the difficulties that the
chiral $p$-wave state has explaining the H-T phase diagram, we
also address the possibility of an in-plane $\vd$-vector to see if
it can account for the observed phase diagram. This follows a
discussion of the limited conditions for which an in-plane
$\vd$-vector is consistent with experimental results. We find that
an in-plane $\vd$-vector that is nearly free to rotate in the
plane can explain the H-T phase diagram.

The paper begins with an overview of Ginzburg Landau theory for
the chiral $p$-wave state to provide the origin of the two
predictions (i) and (ii) above. Then we discus the role of
spin-orbit coupling on the orientation of the $\vd$-vector. This
discussion motivates an examination of the the quasi-classical
equations in a magnetic field for which more than one irreducible
representation is important. The quasi-classical equations are
solved using an approximation that is valid in high field regime
for all temperatures. We present the resulting H-T phase diagrams
for the chiral $p$-wave state. Finally, after a discussion of the
consistency of an in-plane $\vd$-vector with existing experimental
results,  we present results on the H-T phase diagram for
this case as well.  

\section{Ginzburg Landau Theory}
The simplest framework within which the role of magnetic fields on
the chiral $p$-wave state can be understood is the Ginzburg Landau
theory. Here we give a brief overview of this theory to
demonstrate the origin of the additional  transition and the
anisotropy in the upper critical field that we will discuss later
within a microscopic theory. The free energy density for the $E_u$
representation of $D_{4h}$ with a basis $\veta=(\eta_x,\eta_y)$
[this basis has the same rotation properties as $(x,y)$] is given
by \cite{sig91,gor87}
\begin{eqnarray}
f=&-|\veta|^2+|\veta|^4/2+
\beta_2(\eta_x\eta_y^*-\eta_y\eta_x^*)^2/2
+\beta_3|\eta_x|^2|\eta_y|^2 +
|D_x\eta_x|^2+|D_y\eta_y|^2 \nonumber \\
& +\kappa_2(|D_y\eta_x|^2+ |D_x\eta_y|^2)+\kappa_5(|D_z\eta_x|^2+
|D_z\eta_y|^2)\label{eq1}
\\ & +\kappa_3[(D_x\eta_x)(D_y\eta_y)^*+h.c.]+
\kappa_4[(D_y\eta_x)(D_x\eta_y)^*+h.c.] +\vh^2/(8\pi), \nonumber
\end{eqnarray}
where $D_j=\nabla_j-\frac{2ie}{\hbar c} A_j$, $\vh=\vnabla\times
{\bf A}$, and ${\bf A}$ is the vector potential.  There are three
possible homogeneous phases \cite{sig91,gor87}: (a)
$\veta=(1,i)/\sqrt{2}$ ($\beta_2>0$ and $\beta_2>\beta_3/2$), (b)
$\veta=(1,0)$ ($\beta_3>0$ and $\beta_2<\beta_3/2$), and (c)
$\veta=(1,1)/\sqrt{2}$ ($\beta_3<0$ and $\beta_2<0$). Phase (a) is
nodeless (if the Fermi surface has the same topology as a
cylinder) and phases (b) and (c) have line nodes. Weak coupling
theories give rise to phase (a): the chiral $p$-wave phase. The
application of a magnetic field in the basal plane breaks the
degeneracy of the two components $\eta_x$ and $\eta_y$. For the
chiral $p$-wave state, symmetry arguments imply that the vortex
lattice phase diagram contains at least two vortex lattice phases
for magnetic fields applied along any of the four two-fold
symmetry axes: $\{(1,0,0),(0,1,0),(1, 1,0),(1,-1,0)\}$
\cite{agt98,agt00}.
To illustrate the origin of these phase transitions, consider a
zero-field ground state $\veta=(1,i)$ and a magnetic field applied
along the $(1,0,0)$ direction. Due to the broken tetragonal
symmetry, the degeneracy of the $\veta=(1,0)$ and the
$\veta=(0,1)$ solutions is removed by the magnetic field.
Consequently, only one of these two possibilities will order at
the upper critical field. However, if the system is spatially
uniform along the magnetic field, then the solution near the upper
critical field will exhibit a symmetry that the zero-field
solution does not. For our example, this symmetry is either
$\sigma_x$ (if $\veta=(0,1)$ orders at $H_{c2}$)  or
$-\sigma_x=U(\pi)\sigma_x$ (if $\veta=(1,0)$ orders at $H_{c2}$)
where $U(\pi)$ is a gauge transformation and $\sigma_x$ is a
reflection through the $x$-axis. The only way this can occur is if
there is an additional phase transition as magnetic field is
reduced to break this symmetry. The only difference that occurs
for the field applied along the $(1,1,0)$ direction is that the
solution near the upper critical field will be either
$\veta=(1,1)$ or $\veta=(1,-1)$.

Another result of the $E_u$ theory that follows from
Eq.~\ref{eq1}, originally shown by Gor'kov, is that the upper
critical field is anisotropic near $T_c$ \cite{gor87}. Such an
anisotropy, for which $dH_{c2}/dT|_{T=T_c}$ is not equal for
$(1,0,0)$ and the $(1,1,0)$ directions, cannot occur for order
parameters that have only one complex degree of freedom
\cite{gor87}. The anisotropy in upper critical field near $T_c$
has been calculated from microscopic calculations for the in-plane
fields, along the $(1,0,0)$ and $(1,1,0)$ directions, for a gap
function of the form $\vd(\vk)=\mhz[\eta_x f_x(\vk)+\eta_y
f_y(\vk)]$\cite{agt01}. These calculations show that anisotropy is
generally much larger than that experimentally observed. However,
under certain special circumstances, this anisotropy can be small
\cite{agt01}. To examine the lack of anisotropy for the whole
temperature range requires a microscopic model that goes beyond
the Ginzburg Landau theory as is done below. Note that previous
microscopic studies of the chiral $p$-wave state for in-plane
magnetic fields \cite{kus04,uda04} did not show reveal the physics
discussed here. This was because the order parameter in these
works was fixed to have the form $\veta=(1, i)$ for all magnetic
fields and temperatures. Such an approximation is valid only for
fields much smaller than $H_{c2}$.

\section{Spin-orbit coupling and the orientation of $\vd$}

An important interaction in determining the specific spin-triplet
pairing state in Sr$_2$RuO$_4$ is spin-orbit coupling. We quantify
this in this section. From a symmetry point of view, the
superconducting state belongs to one of the odd-parity
representations of the tetragonal point group
$D_{4h}$\cite{sig91,gor87}. The quasi two-dimensionality of the
Fermi liquid state in Sr$_2$RuO$_4$ makes it reasonable to assume
that the momentum dependence of the superconducting state is
described by functions $f_x(\vk)$ and $f_y(\vk)$ which obey the
same symmetry transformation properties as $k_x$ and $k_y$
respectively under rotations of $D_{4h}$ (but otherwise are
arbitrary). When there is no spin-orbit coupling, the spin-triplet
state has six-fold degeneracy \cite{Rice,Ng87}, the transition
temperature $T_c$ will be same for any linear combination for gap
functions given in Table 1. When spin-orbit coupling is included,
the degeneracy of $T_c$ will be lifted. The stable state will
either have the $\vd$-vector along c-axis or be a linear
combination of in-plane $\vd$-vectors.

To quantify the role of spin-orbit coupling, we follow an approach
developed by Sigrist {\it et al.} \cite{sig02}. In particular, the
effect of spin-orbit coupling is included through the magnetic
susceptibility. The Hamiltonian with a general pairing interaction
is
\begin{equation} \begin{array}{l}
{\cal H} = \displaystyle\sum_{\vsk,s}\epsilon_\vk c^{\dag}_{\vsk s}
c_{-\vsk s}+\frac{1}{2} \sum_{\vsk, \vsk'} \sum_{s_1 s_2 s_3 s_4}
V_{\vk,\vk';s_1 s_2 s_3 s_4}
c^{\dag}_{\vsk s_1} c^{\dag}_{-\vsk s_2} c_{-\vsk' s_3}
                     c_{\vsk' s_4},
                     \end{array}
\end{equation}
where $\epsilon_\vk$ is electron band energy measured from the
Fermi energy and $c^{\dag}_{\vsk s}, c_{\vsk s}$ are the fermion
creation and annihilation operators. As a concrete model, we use
an effective pairing interaction that is due to spin
fluctuations\cite{sig02}. However, the results that we require
later depend solely upon the splitting of the six-fold degeneracy
(this can be incorporated in a model independent way within the
quasi-classical theory). The effective pairing interaction we use
is
\begin{equation}
V_{\vsk \vsk' ,s_1,s_2,s_3,s_4} = -{I^2\over 16} \sum_{\mu,\nu}\{
\left[\chi_{\mu,\nu}(\vsk,\vsk')+
\chi_{\nu,\mu}(\vsk,\vsk')\right]\sigma^\mu_{s_1,s_4}\sigma^\nu_{s_2,s_3}-
\left[\chi_{\mu,\nu}(-\vsk,\vsk')+
\chi_{\nu,\mu}(\vsk,-\vsk')\right]\sigma^\mu_{s_2,s_4}\sigma^\nu_{s_1,s_3}\},
\end{equation}
where $I$ is an coupling constant, and $\chi_{\mu \nu}(\vk,\vk')$
is the static susceptibility. The phenomenological form of
$\chi_{\mu,\nu}(\vk,\vk')$ in a material of tetragonal symmetry is
\begin{equation}
\chi_{\mu,\nu}(\vk,\vk')=\left(\begin{array}{ccc}
g_1(f_x f'_x+f_y f'_y)+g_2(f_x f'_x-f_y f'_y) & g_3(f_x f'_y+f_y f'_x) & 0\\
 g_3(f_x f'_y+f_y f'_x) & g_1(f_x f'_x+f_y f'_y)-g_2(f_x f'_x-f_y f'_y) & 0\\
 0 & 0 & g_z(f_x f'_x+f_y f'_y)\end{array}\right)\label{suscep}, \end{equation}
where $g_1, g_2, g_3$ and $g_z$ are phenomenological parameters. The
self consistency equation for $\vd$-vector with the above
interaction can be solved to get the superconducting transition
temperature, $k_B T_c=1.14 \omega_c exp(-16/I^2 N(0) V_{\Gamma})$
for the different representations $\Gamma$ ($V_{\Gamma}$ corresponds
to the interaction for the representation $\Gamma$). These are
listed in Table I.
\begin{table}[ht]
\begin{center}
\begin{tabular}{|c|c|c|}\hline
Rep $(\Gamma)$ & Gap Function  & Interaction ($V_\Gamma$) \\\hline
\hline
 $A_{1u}$ & $\hat{x} f_x+\hat{y} f_y$ & $g_z-2(g_2+g_3)$  \\\hline
 $A_{2u}$ & $\hat{x} f_y-\hat{y} f_x$ & $g_z+2(g_2+g_3)$  \\\hline
 $B_{1u}$ & $\hat{x} f_x-\hat{y} f_y$ & $g_z-2(g_2-g_3)$  \\\hline
 $B_{2u}$ & $\hat{x} f_y+\hat{y} f_x$ & $g_z+2(g_2-g_3)$  \\\hline
 $E_{u}$  & $\hat{z}(f_x\pm i f_y)$   & $2g_1-g_z $       \\\hline
 \end{tabular}
\end{center}
\caption{Gap functions and interaction strengths for the different
representations of $D_{4h}$.} \label{table}
\end{table}
In the limit $g_1=g_z$ and $g_2=g_3=0$ there is no spin orbit
coupling and all the representations will be degenerate. If there
is cylindrical symmetry then $g_2=g_3$. Notice that this does not
imply that that all the in-plane pairing states are degenerate.
For this to occur, $g_2=g_3=0$. It is instructive to use results
from recent microscopic calculations to gain an insight into the
relative size of $\{g_1,g_2,g_3,g_z\}$ \cite{Ng87,yan03}. Both
these papers reveal that deviation from the isotropic limit is
small, since all the representations have very similar transition
temperatures. The results of Ref.~\onlinecite{yan03} correspond to
the limit $g_3=0$, $g_1>g_z$, and $|g_1-g_z|>>|g_2|$. Based on
these results we will assume that $|g_2|,|g_3|<<|g_1-g_z|$. The
case $g_1>g_z$ corresponds to the chiral pairing state
\cite{Ng87,sig02}. While the case $g_z>g_1$ and
$|g_2|,|g_3|<<|g_1-g_z|$ corresponds to the nearly degenerate
in-plane $\vd$-vector. We will consider both these cases in the
following.





\section{Eilenberger equations for the $\gamma$ band}

An important aspect for understanding the superconducting state in
Sr$_2$RuO$_4$ is the band structure. In particular, the states
near the Fermi surface are derived from the Ru $t_{2g}$ orbitals.
The degeneracy of these orbitals are split by the tetragonal
crystal field into a $xy$ orbital and the degenerate $\{xz,yz\}$
orbitals \cite{ogu95,sin95,mac96,ber00}. These two sets of
orbitals have a different parity under a mirror reflection through
the basal plane. Consequently, to first approximation, the
$\gamma$ sheet of the Fermi surface is comprised of $xy$ Wannier
functions, while the $\{\alpha,\beta\}$ sheets of the Fermi
surface are comprised of $\{xz,yz\}$ Wannier functions. This leads
naturally to orbital dependent superconductivity
\cite{agt97,zhi01}; a theory for the superconducting state has
different gaps on the $\{\alpha,\beta\}$ and $\gamma$ sheets of
the Fermi surface. This theory has experimental support through
specific heat measurements in magnetic fields
\cite{deg04,deg04-2}. These measurements indicate that for strong
fields applied in the basal plane, superconductivity in the
$\{\alpha,\beta\}$ bands is suppressed and the $\gamma$ band has
the dominant superconducting gap. In addition to these
measurements, recent theoretical calculations indicate that the
ratio of the $\{\alpha,\beta\}$ band gaps to that of the $\gamma$
band gap is $\approx 0.15$ in the high field limit \cite{kus04}.
Therefore, we restrict the following microscopic theory in the
high field regime to a single band theory for the $\gamma$ band.


Now we will explain briefly the approximate analytic solution of
the fundamental quasi-classical equations for a single band within
weak-coupling superconductivity under magnetic fields. Our
notation and formulation  follows that of Ref.~\onlinecite{kus04}.
The solution can be obtained by the following approximations: (i)
the spatial dependence of the internal magnetic field is averaged
by $B$, (ii) the vortex lattice structure is expressed by the
Abrikosov lattice and (iii) the diagonal elements of the Green's
function are approximated by the spatial average. In general, a
magnetic field will mix different representations of the $D_{4h}$.
Consequently, the $\vd$-vector will be a linear combination of the
functions listed in Table I,
\begin{eqnarray}
\vd(\vR,\vk)=&d_1(\vR)[\hat{x} f_x(\vk)+\hat{y}
f_y(\vk)]+d_2(\vR)[\hat{x} f_y(\vk)+\hat{y} f_x(\vk)]
+d_3(\vR)[\hat{x} f_x(\vk)-\hat{y} f_y(\vk)] \nonumber
\\&+d_4(\vR)[\hat{x} f_y(\vk)-\hat{y} f_x(\vk)] +
\hat{z}[\eta_x(\vR) f_x(\vk)+ \eta_y (\vR)f_y(\vk)],
\label{dwhole}
\end{eqnarray}
the form of $f_x(\vk)$ and $f_y(\vk)$ is given in next subsection.
The approximation (ii) amounts to taking each order parameter
component in the lowest Landau level, so that $\vd(\vR,\vk)=\Delta
\phi_0 (\vR) \bf{\tilde{d}}(\vk)$, where $\phi_0$, the lowest
Landau level, is given by
\begin{equation} \phi_0(\vR)=\sum_n c_n e^{-ip_n
y'}\exp[-((x'-\Lambda^2p_n)/\Lambda)^2/2],
\end{equation}
where $p_n=2\pi n/\beta$, $\beta$ being the lattice constant in
the $y$ direction, $\Lambda=(2|e|B)^{-1/2}$ is the magnetic
length, 
and the coefficients $c_n$, which determine the type of vortex
lattice, are such that $<|\phi_0(\vR)|^2>=1$, $\Delta$ is the
magnitude of gap, and ${\bf \tilde{d}}(\vk)$ defines the angular
dependence of $\vd$-vector. We have taken anisotropy into account by
writing $x=x'/\chi^{1/2}$ and $y=y'\chi^{1/2}$. For a conventional
superconductor, even though all approximations mentioned above are
valid near $H_{c2}$, comparisons with reliable numerical
calculations suggest that the solution is competent quantitatively
in wide region of the $(T,H)$ phase diagram except in very low $T$
and $H$ regions \cite{kus04-2}. For further detail we recommend the
reader to refer to the literature
\cite{Brandt67,Pesch75,Houghton98,Dahm02,kus04-2}. In principle, we
should consider higher Landau level solutions in this problem.
However, within Ginzburg Landau theory our solution is exact in the
high field limit provided the magnetic field is in the basal plane.
This indicates that it is reasonable to keep only the lowest Landau
level (for the field along the $c$-axis, other Landau levels must be
included \cite{agt98}). The expression for the free energy measured
relative to the normal state energy \cite{kus04-2}, which is given
for strongly type-II superconductors, $B\simeq H$, in the clean
limit is,
\begin{eqnarray}
\Omega_{\rm SN}/N_0= \ln\left({T\over
T_c^z}\right)(|\eta_x|^2+|\eta_y|^2)+
\sum_{\Gamma=1}^{4}|d_\Gamma|^2 \ln\left({T\over T_c^\Gamma}\right)
+2\pi T \sum_{n=0}^\infty\left(\frac{\Delta^2}{\omega_n}-\langle I
\rangle \right), \label{free}
\end{eqnarray}
where $N_0$ is the total density of states (DOS) in the normal
state, and $\omega_n=(2n+1)\pi T$ is the Fermionic Matsubara
frequency. Eq.~\ref{free} generalizes the corresponding
expressions in earlier works
\cite{Brandt67,Pesch75,Houghton98,Dahm02,kus04-2} by including
more than one irreducible representation. The function $I$ is
given by
\begin{equation}
I=\frac{2g}{1+g}\sqrt{\pi}\left(\frac{2\Lambda}{\tilde{v}_{\perp}(\hvk)}\right)\Delta^2|\varphi(\hvk)|^2W(iu_{
n}),
\end{equation}
with
\begin{equation}
g=\left[
1+\frac{\sqrt{\pi}}{i}\left(\frac{2\Lambda}{\tilde{v}_{\perp}(\hvk)}\right)^2\Delta^2|\varphi(\hvk)|^2W'(iu_{
n}), \right]^{-1/2},
\end{equation}
where $u_{ n}=2\Lambda \omega_n/\tilde{v}_{\perp}(\vk)$,
$W(z)=e^{-z^2}{\rm erfc}(-iz)$ is the Faddeeva function, and
$|\varphi(\hvk)|^2={\bf \tilde{d}}(\vk)~\cdot~{\bf
\tilde{d}}^*(\vk)$. Here $\tilde{v}_{\perp}(\hvk)$ is the
component of $\vv$ perpendicular to the field, which for an
in-plane field of the form ${\vH}=H(\cos\phi_h,\sin\phi_h,0)$, is
given as
\begin{equation}
\tilde{v}_{\perp}^2(\hvk)=\chi^{-1/2}v_z^2+\chi^{1/2}v_F^2\sin^2(\phi-\phi_h),
\label{vperp}
\end{equation}
here $\chi=\tilde{\chi}v_F/v_c$ is an anisotropy parameter (we let
$\tilde{\chi}$ be arbitrary). We have taken $v_z(k)=v_c
\textmd{sgn}(k_z)$. All fields are measured in units
$2\pi^2T_c^2/(e v_Fv_c)$ and $\hbar$ and $k_B$ are both 1.
We will use this formalism to examine the in-plane magnetic field
phase diagram for the chiral $p$-wave state and the nearly
degenerate in-plane $\vd$-vector separately.

\subsection{Momentum dependence of the gap function}

To complete the description of the superconducting state, we must
specify the functions $f_i(\vk)$ ($i=\{x,y\}$).  The most general
gap function consistent with translational invariance and with the
appropriate rotational properties of $D_{4h}$ is
\begin{equation}
f_i(\vk)=\sum_{n=1,m=0}^\infty c_{n,m}\sin(n k_i)\cos(m k_j),~~~~~
i,j=x,y~~ i\neq j,
\end{equation}
where $c_{n,m}$ are complex coefficients, $k_i$ in units $\pi/a$,
and $a$ is the lattice spacing. Here $n=1,~m=0$ represents a
Cooper pair formed by nearest neighbor (nn) interactions and
$n=1,~m=1$ to a Cooper pair formed from next nearest neighbor
(nnn) interactions. In general, increasing $m,n$ corresponds to
forming Cooper pairs from interactions between increasing number
of neighbors. We will restrict ourselves here to nn and nnn
pairing interactions. This has some support from microscopic
calculations. In particular, the theory for nn interactions was
originally proposed by Miyake and Narikiyo \cite{miy99} and was
also examined by Nomura and Yamada \cite{nom00}. Cooper pairs for
which nn interactions are not important but which have a
substantial contribution from nnn interactions have been proposed
recently by Arita {\it et al.} \cite{ari04} with the assumption of
large on-site and nearest neighbor Coulomb interactions. In
particular, the latter paper proposes a gap function of the form
$\vd=\mhz[\sin(k_x+k_y)-i\sin(k_x-k_y)]=\mhz
\sqrt{2}e^{-i\pi/4}[\sin k_x \cos k_y+i\sin k_y \cos k_x])$. We
take Cooper pairs to be formed by nn and nnn interactions and take
the form of $f_x$ to be, $f_x=\sin(k_x)[1+\epsilon \cos(k_y)]$. We
keep the parameter $\epsilon$ to be arbitrary and allow it to
vary.  We take the Fermi surface to be cylindrical with
$(k_x,k_y)=\pi R (\cos \phi,\sin \phi)$ where $R=0.9$ approximates
the Fermi surface of the gamma band (in the third subsection of
next section we will take R=0.79 for reasons that will be
apparent).

\section{Chiral $p$-wave state}
For the chiral $p$-wave state, we use
$\vd(\vk,\vR)=\mhz\Delta\phi_0(\vR)\varphi(\hvk)$ with
\begin{equation}
\varphi(\hvk)=[\cos\theta f_x(\vk) +\sin\theta
e^{i\zeta}f_y(\vk)], \label{dz}
\end{equation}
(not normalized for notational simplicity). For this gap function,
the Ginzburg Landau order parameter takes the specific form
$\veta(\vR)=\Delta\phi_0(\vR)(\cos\theta,e^{i\zeta}\sin\theta)$.

\subsection{Upper critical field}
\begin{figure}[ht]
\epsfxsize=3.5 in \center{\epsfbox{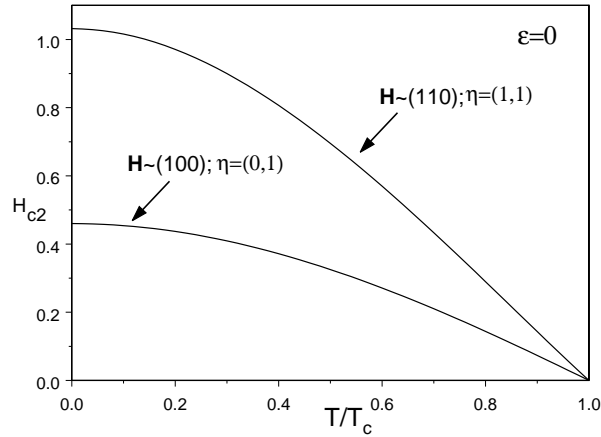}} \caption{Upper
critical fields for $\epsilon=0$ for the fields along the (1,0,0)
and (1,1,0) directions. The field in units $2\pi^2T_c^2/(e v_Fv_c)$
} \label{fig1}
\end{figure}

\begin{figure}[ht]
\epsfxsize=3.5 in \center{\epsfbox{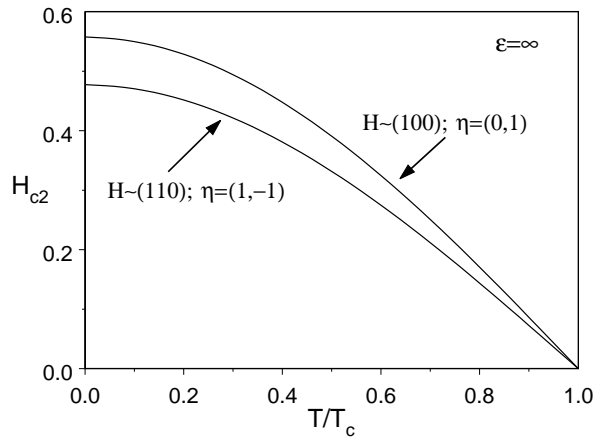}} \caption{Upper
critical fields for $\epsilon=\infty$ and for the fields along the
(1,0,0) and (1,1,0) directions.} \label{fig2}
\end{figure}

\begin{figure}[ht]
\epsfxsize=3.5 in \center{\epsfbox{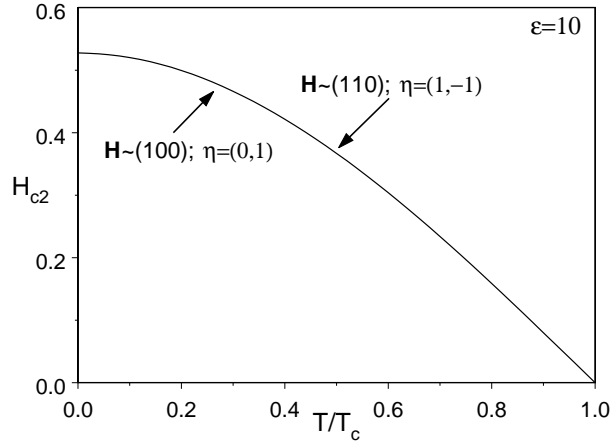}} \caption{Upper
critical fields for $\epsilon=10$ and for the fields along the
(1,0,0) and (1,1,0) directions. The two upper critical fields
almost identical.} \label{fig3}
\end{figure}

The anisotropy in upper critical field for (1,0,0) and (1,1,0)
directions is a generic feature of the $E_u$ theory. Since
experimentally it has been observed that the upper critical field
is relatively isotropic for in-plane fields \cite{mao00}, we ask
if it is possible to reproduce this. The only free parameter in
the theory is $\epsilon$ which describes the anisotropy in the gap
function. Surprisingly, we have found that it is possible for
$\epsilon=10$. The upper critical field for the field in the
$(1,1,0)$ and the $(1,0,0)$ directions for the three values of
$\epsilon$; $\epsilon=0$, $\epsilon=\infty$, and $\epsilon=10$ for
the chiral superconducting state has been shown in
Figs.~\ref{fig1}, \ref{fig2}, and \ref{fig3}. 
In these figures we also show the stable solutions for the order
parameter at $H_{c2}$. For a field along the $(1,0,0)$ direction,
the possible stable solutions are $\veta=(1,0)$ or $\veta=(0,1)$.
For a field along the $(1,1,0)$ direction, the possible stable
solutions are $\veta=(1,1)$ or $\veta=(1,-1)$.

\subsection{Phase diagram}
\begin{figure}[ht]
\epsfxsize=3.5 in \center{\epsfbox{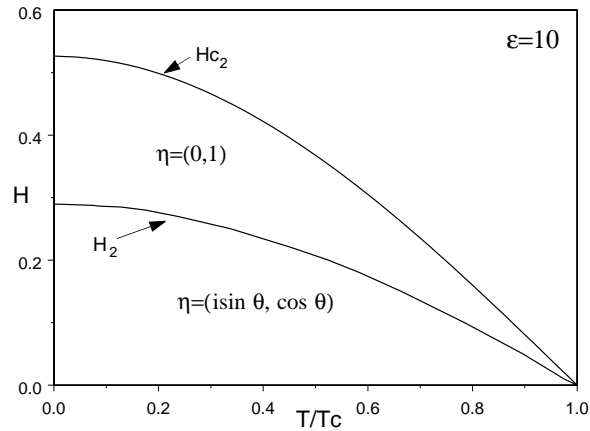}} \caption{Phase
boundaries for $\epsilon=10$ and for the field along the $(100)$
direction.} \label{fig4}
\end{figure}

Another generic feature of the $E_u$ theory is that multiple
vortex phases exist for in-plane fields. For a field along the
$(1,0,0)$ direction, the solution near $H_{c2}$ is $\veta=(0,1)$
(for the three values of $\epsilon$ discussed above this was the
case) then as field is reduced for fixed temperature, a second
transition occurs at $H_2$ for which the $\veta=(1,0)$ component
becomes non-zero. Such phase transitions have only been examined
within Ginzburg Landau theory \cite{agt97,kit99}. The Eilenberger
equations discussed above allow for the examination of this phase
transition throughout the entire H-T phase diagram. Here, we apply
this approach to the $\epsilon=10$ gap function for a $(1,0,0)$
field direction, the resulting phase diagram is shown in
Fig.~\ref{fig4}.

\begin{figure}[ht]
\epsfxsize=3.5 in \center{\epsfbox{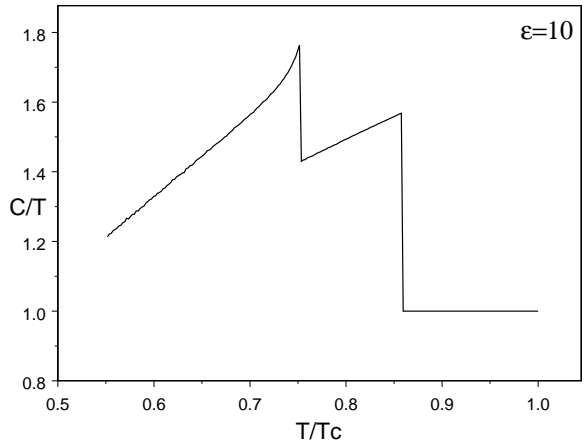}} \caption{Specific
heat as a function of temperature for a fixed field
$H/H^0_{c2}=0.21$ and for $\epsilon=10$.} \label{fig6}
\end{figure}

The specific heat as a function of temperature is also shown in
Fig.~\ref{fig6} for $H/H^0_{c2}=0.21$. This plot clearly shows a
second transition in the specific heat. Such a second transition
has not been seen in specific heat measurements. This represents a
difficulty for the $E_u$ theory. In general, we have not been able
to find a microscopic theory that can account for both the lack of
anisotropy in the upper critical field and the lack of the second
transition. It is possible that experiments have not seen the
predicted specific heat jumps due to the broadening associated
with fluctuations in the vortex phase, or due to sample
inhomogeneities. In the case of UPt$_3$, for which multiple phase
transitions in the vortex phase have been observed, the entire
phase diagram was found through ultrasound measurements
\cite{ade90}. Specific heat measurements mapped out some portions
of the phase diagram \cite{fis89} but they did not show clear
anomalies throughout the entire phase diagram \cite{ram95}.
Therefore, it would be useful to look for such transitions in the
phase diagram for Sr$_2$RuO$_4$ with other probes such as
ultrasound.


\subsection{Tetracritical point}

While there has been no evidence of multiple superconducting
transitions in the low-field regime, two superconducting
transitions have been observed in the high-field range
\cite{mao00,mao00-2}. In one aspect these transitions are natural
candidates for two transitions discussed above. In particular, the
vanishing of the second transition (which does not occur at
$H_{c2}$) as the field is rotated away from the in-plane direction
is consistent with the above predictions. However, the second
transition is only observed for $T/T_c<0.1$ and appears to
intersect the upper critical field line. This is inconsistent with
the above prediction which predicts that this transition should
exist for all temperatures. Here we explore a possible explanation
for this transition that is based on results of the Eilenberger
equations.

We have found that for small parameter ranges in the model
described above, it is possible that the solution for the order
parameter at the upper critical field {\it changes} as a function
of temperature. In particular for a field along the $(1,0,0)$
direction, the low temperature solution (at $H_{c2}$) is
$\veta=(1,0)$ and there is a transition as temperature is reduced
so that the solution at $H_{c2}$ becomes $\veta=(0,1)$. The phase
diagram that emerges bears a striking similarity to the observed
results.

\begin{figure}[ht]
\epsfxsize=4.0 in \center{\epsfbox{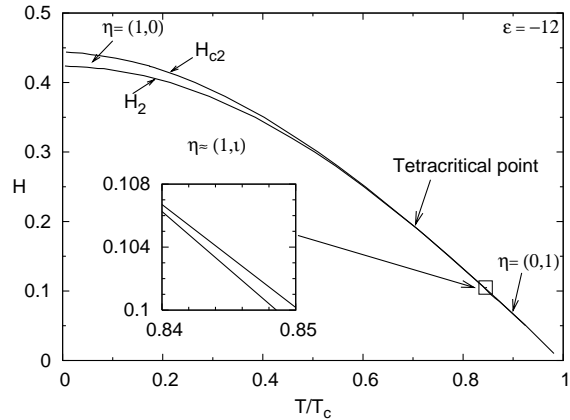}} \caption{Phase
diagram showing a tetracritical point. Between $T_c$ and the
tetracritical  temperature there are two phase transitions, shown
in the inset, that are difficult to distinguish from each other.}
\label{tetra phase}
\end{figure}

\begin{figure}[ht]
\epsfxsize=3.5 in \center{\epsfbox{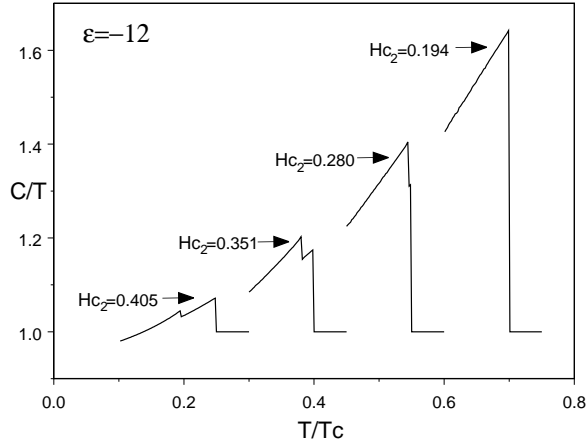}} \caption{Specific
heat as a function of temperature for fixed magnetic fields.}
\label{tetra sp}
\end{figure}

In Figs.~\ref{tetra phase} and \ref{tetra sp}, we show the phase
diagram and specific heat calculations for $R=0.79$ and
$\epsilon=-12$. Note that a negative $\epsilon$ corresponds to a
repulsive interaction between Cooper pairs formed from nearest and
Cooper pairs formed from next nearest neighbors. This theory would
still require the existence of two transitions up to $T=T_c$.
However, as this phase diagram shows, the two transition lines
between $T_c$ and the temperature of the tetracritical point are
very close to each other and will be very difficult to observed
experimentally.

While this phase diagram agrees with that observed experimentally
for a field along $(1,0,0)$ direction, this choice of parameters
also exhibits a moderate $\approx 15\%$ anisotropy of the in-plane
upper critical field. Furthermore, for the field along the
$(1,1,0)$ direction the phase diagram resembles that of Fig.~4
(there is no tetracritical point). Therefore, these results can
only be taken as suggestive since this set of parameters cannot
account for all observed features. It is possible that the gap on
the $\{\alpha,\beta\}$ bands may improve the agreement between
theory and experiment.  It appears that at low fields these gaps
cannot be neglected \cite{kus04} and the suppression of these gaps
relative to that of the $\gamma$ may provide a more robust
mechanism for the appearance of a tetracritical point in the
relatively low field regime. This can occur if these bands prefer
orthogonal order parameter solutions at $H_{c2}$, as was often the
case for calculations for different $E_u$ gap functions in
Ref.~\onlinecite{agt01}. A second possible explanation for both
the observed lack of anisotropy and the existence of a
tetracritical point is that the parameter $\epsilon$ may not be
constant for all temperatures and magnetic fields as we have used
here. A complete description would require an effective two-gap
theory for which $\epsilon$ is determined self-consistently. In
such a theory, $\epsilon$ can develop a temperature and magnetic
field dependence (though in many circumstances, $\epsilon$ will be
approximately constant). A complete  analysis of this would
require a detailed knowledge of the microscopic interactions
giving rise to superconductivity. This is beyond the scope of this
paper.



\section{Nearly degenerate in plane $\vd$-vector}

The chiral $p$-wave state has difficulties explaining both the
observed absence of anisotropy in $H_{c2}$ and the absence of
additional phase transitions in the vortex phase. For this reason
we also consider the nearly degenerate in-plane $\vd$-vector.
Initially, we consider under what circumstances the nearly
degenerate $\vd$-vector is consistent with other experiments
(Knight shift, $\mu$SR, vortex lattice structure, and Josephson
experiments). Then we examine the specific heat as a function of
magnetic field in the vortex state and show that the second
anomaly is rapidly suppressed by the breaking of the degeneracy of
the four components of the in-plane $\vd$-vector.

Knight shift measurements can be naturally accounted for by an
in-plane $\vd$-vector. The most recent observation is that for
magnetic fields along $c$-axis there is no change in the spin
susceptibility as temperature is reduced \cite{mur04}. The
simplest interpretation of this result is that the $\vd$-vector is
in the basal plane. The earlier observation that the spin
susceptibility is unchanged for in-plane magnetic fields would
require that the $\vd$-vector is free to rotate in the basal
plane. This implies that all the in-plane $\vd$-vector states are
degenerate or nearly degenerate. The observed square vortex
lattice for the field along the $c$-axis is also consistent with a
degenerate in-plane $\vd$-vector. This will follows from an
Ginzburg Landau analysis, where it can be shown that free energy
for the degenerate in-plane $\vd$-vector has equilibrium solutions
with the same properties as those of the Ginzburg Landau theory
for the chiral $p$-wave state.

The muon spin relaxation ($\mu$SR) measurements of Luke {\it et
al.} \cite{luk98} have found an increased spin relaxation rate in
the superconducting state with zero applied magnetic field. This
has commonly been interpreted as evidence for a superconductor
that breaks time reversal symmetry in the bulk. However, any bulk
internal magnetic field must be screened due to the Meissner
effect. Consequently, $\mu$SR only probes internal magnetic fields
due to inhomogeneities such as impurities or domain walls between
degenerate superconducting states. It has been shown that a
superconducting state that does not break time reversal symmetry
in the bulk can still give rise to local internal fields
\cite{sig91}. The important condition for such internal magnetic
fields to exist is that the superconducting order parameter has
more than one degree of freedom. This distinction is emphasized
here because for nearly degenerate in-plane $\vd$-vectors, the
bulk superconducting state in zero applied field does not break
time reversal symmetry. This does not imply that such a state is
inconsistent with $\mu$SR measurements. However, it does require
that the different in-plane $\vd$-vector representations are
nearly degenerate (the different $T_c$ values must lie close to
each other).

The most difficult experiments to explain with an in-plane
$\vd$-vector are the Josephson experiments. The most recent of
these has found that for Sr$_2$RuO$_4$-Au$_{0.5}$In$_{0.5}$ SQUID,
there is a $\pi$ phase difference in the Josephson current when
the two junctions have  opposite normals \cite{nel04}. While this
is generally expected for a $p$-wave superconductor, it cannot be
explained by an in-plane $\vd$-vector.  Such a $\vd$-vector does
not allow for a Josephson current between an odd-parity
superconductor and an isotropic ($s$-wave) superconductor when the
junction has a normal perpendicular to the Sr$_2$RuO$_4$ $c$-axis.
The existence of such a Josephson current implies a $\vd$-vector
aligned along the $c$-axis. An explanation of such a Josephson
current within an in-plane $\vd$-vector approach would therefore
require that at an interface, the $ \vd$-vector is along the
$c$-axis and in the bulk it is in-plane. A similar scenario has
been proposed by Bahcall in the context of the cuprate
superconductors (in this case, the order parameter near the
interface is $s$-wave and becomes $d$-wave in the bulk)
\cite{Bah96}. In support of such a picture, the $\vd$-vector is
almost certainly along the $c$-axis for an interface with a normal
perpendicular to the $c$-axis. This is a natural consequence of a
stronger spin-orbit coupling at the interface than in the bulk
(spin-orbit coupling is governed by the gradient of the
single-particle potential). It is the spin-orbit coupling that
governs the orientation of the $\vd$-vector. A Rashba spin-orbit
coupling of the form $\alpha_R\hat{n}\cdot \vk\times\vS(k)$ (where
$\hat{n}$ is the interface normal, $\vk$ is the fermion
wave-number, $\vS(k)$ is the fermion spin, and $\alpha_R$ is a
coupling constant) would give a $\vd$ with a component along the
$c$-axis if $\hat{n}$ lies perpendicular to the $c$-axis
\cite{fri04}. If the bulk $\vd$-vector is in-plane and the
$\vd$-vector lies along the $c$-axis near the interface, then an
analysis following that of Bahcall would imply that the $\pi$
squid experiment of Nelson {\it et al.} should sometimes see a
$\pi$ phase shift and sometimes no phase shift.  Nelson's data
indicates that there is always a $\pi$ phase shift. However, given
that data on only three samples are presented, it may be prudent
to await further results before ruling out an in-plane
$\vd$-vector in the bulk on the basis of these experiments.

The general form for an  in-plane $\vd$-vector will be a linear
combination of four different in-plane representations listed in
Table I,
\begin{eqnarray}
\vd(\vR,\vk)=&d_1(\vR)[\hat{x} f_x(\vk)+\hat{y}
f_y(\vk)]+d_2(\vR)[\hat{x} f_y(\vk)+\hat{y} f_x(\vk)]
+d_3(\vR)[\hat{x} f_x(\vk)-\hat{y} f_y(\vk)] \nonumber
\\&+d_4(\vR)[\hat{x} f_y(\vk)-\hat{y} f_x(\vk)]. \label{dwhole}
\end{eqnarray}
We parameterize $[d_1(\vR),d_2(\vR),d_3(\vR),d_4(\vR)]$ as
$\Delta\phi_0(\vR) (\cos\psi \cos\theta,\sin\psi \sin\phi,
\cos\psi \sin\theta,\sin\psi \cos\phi )$ so that $\vd$-vector
 can be written as
\begin{equation}
\vd(\vR,\vk)=\Delta(\vR)[\hat{x}\varphi_x(\vk)+
\hat{y}\varphi_y(\vk)] \label{dinplane},
\end{equation}
with
$\varphi_x(\hvk)=\cos\psi(\cos\theta+\sin\theta)f_x(\vk)+\sin\psi(\sin\phi+\cos\phi)f_y(\vk)$
and
$\varphi_y(\hvk)=\cos\psi(\cos\theta-\sin\theta)f_y(\vk)+\sin\psi(\sin\phi-\cos\phi)f_x(\vk)$.
With these definitions
$|\varphi(\hvk)|^2=|\varphi_x(\hvk)|^2+|\varphi_y(\hvk)|^2$.

Prior to presenting the results for the nearly-degenerate in-plane
$\vd$-vector, we briefly give the results for the degenerate
in-plane $\vd$-vector. This corresponds to the situation that
$g_2=g_3=0$ in Table I. In zero applied field there are many
degenerate solutions for this case. In particular, all the
solutions in Table I and the solutions $\vd(\vk)=\hat{e}(k_x\pm
ik_y)$, where $\hat{e}$ is any unit vector in the basal plane, are
all degenerate ground state solutions \cite{Rice}. If a magnetic
field is applied along any of the two-fold symmetry axes, then
there will be two transitions as field is reduced (as there was in
the chiral $p$-wave case). Unlike the chiral $p$-wave case the
second transition can occur in two ways. To illustrate this,
consider an applied field along $(0,1,0)$ direction, the high
field state (the state corresponding to the transition from normal
state to superconducting state) will be either $\hat{x} k_x$ or
$\hat{x} k_y$(it has been assumed here that the $\vd$-vector
prefers to be perpendicular to the magnetic field). Consider
$\hat{x} k_y$ to be concrete. The second transition will appear as
magnetic field is reduced for fixed temperature. The second
transition exists because of the appearance of either a $\hat{x}
k_x$ component [the corresponding zero-field ground state will be
$\hat{x} (k_x\pm i k_y)$]; or a $\hat{y} k_x$ component (the
corresponding zero-field ground state will then be $\hat{x} k_y\pm
\hat{y} k_x$). Strictly speaking, the latter transition will be
energetically less favorable because the $\vd$-vector is not
perpendicular to the magnetic field.  However, it is the latter
transition that will play a more important role when the
degeneracy between the four in-plane $\vd$-vectors is broken. In
this case the solutions at zero-field belong to a single
irreducible representation while the other zero-field solutions
$\hat{x} (k_x\pm i k_y)$ belong to a mixture of more than one
irreducible representation.

\begin{figure}[ht] \epsfxsize=3.5 in
\center{\epsfbox{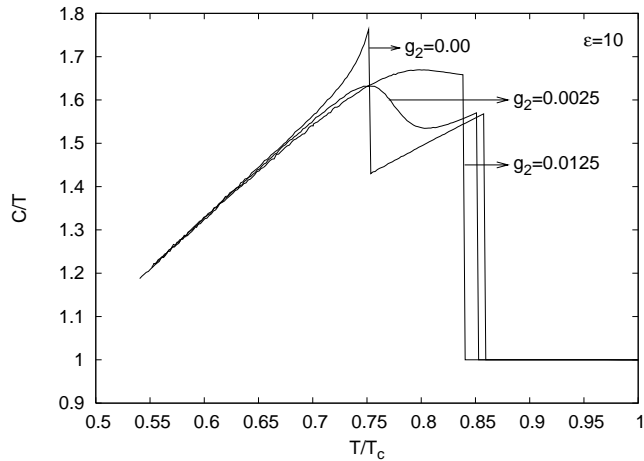}} \caption{Specific heat for
$\epsilon=10$ for the field along the (1,0,0) direction, with
fixed $H/H_{c2}^0=0.21$  for different values of $g_2$ (which is
measured here in units $g_z$).} \label{pert}
\end{figure}

In Fig.~\ref{pert}, we show the specific heat as a function of
temperature for different values of $g_2$ (we have set $g_2=g_3$
in the following) with fixed magnetic field, $H/H_{c2}^0=0.21$
along (0,1,0) direction for $\epsilon=10$ and $R=0.9$. For
$g_2=0$, as discussed above, the second transition will exist and
there are two specific heat anomalies.  The second transition is
removed by a finite value of $g_2/g_z$. The key result is that the
anomaly for the second transition is very quickly suppressed by a
non-zero $g_2/g_z$. Note that the anisotropy in $H_{c2}$ will
still be small for small values of $g_2$. Consequently, a nearly
degenerate in-plane $\vd$-vector can explain the existing
experimental observations on the H-T phase diagram.

The nearly degenerate in-plane $\vd$-vector can account for the
in-plane phase diagram and can qualitatively account for other key
experimental results in Sr$_2$RuO$_4$. However, prior to carrying
out further calculations with this state we note that it should be
possible to rule such a state out experimentally in the near
future. In particular, there are two predictions that can be made
about an in-plane $\vd$-vector. The first has been mentioned
above: further $\pi$ SQUID experiments should reveal the existence
of squids with no phase shift as well as squids with $\pi$ phase
shifts. Also, further Knight shift  experiments should show a
suppression in the spin susceptibility for low enough in-plane
magnetic fields. This will occur because in zero-field the
$\vd$-vector will correspond to a single-component irreducible
representation once $g_2$ and $g_3$ are non-zero and therefore
contain a component that is along the applied field.

\section{Conclusions}

To address an apparent conflict between theoretical predictions of
chiral $p$-wave (the $E_u$ representation) theories of the
superconducting state in Sr$_2$RuO$_4$ and the lack of
corresponding observations, we have carried out quasi-classical
calculations of the superconducting phase diagram for in-plane
magnetic fields.  This has been done for both the chiral $p$-wave
state and for the nearly degenerate in-plane $\vd$-vector. For a
gap function with momentum dependence due to a combination of
nearest and next nearest neighbor interactions defined on the
$\gamma$ band, we find that a small anisotropy in the upper
critical field as the field is rotated in plane is possible.
However, the same gap functions give rise to an additional phase
transition in the vortex state which has not been observed
experimentally. For a narrow range of parameters, the theory gives
rise to a tetracritical point in the H-T phase diagram. When this
tetracritical point exists, the resulting phase diagram closely
resembles the experimentally measured phase diagram for which two
transitions are only observed in the high field regime. We have
also argued that an in-plane $\vd$-vector that can easily rotate
in the basal plane is consistent with existing experimental
results.

\section{Acknowledgements}

We thank Manfred Sigrist for very useful discussions. DFA and RPK
were supported by the National Science Foundation grant No.
DMR-0381665, the American Chemical Society Petroleum Research
Fund. This work was also supported by the Swiss National Science
Foundation.

\end{document}